\documentclass{Interspeech2024}

\interspeechcameraready 
\usepackage{multirow}
\usepackage{algorithm}
\usepackage{algorithmic}
\usepackage{subfigure}
\usepackage{tabularx}

\title{Residual Speaker Representation for One-Shot Voice Conversion}

\name[affiliation={1,2}]{Le}{Xu}
\name[affiliation={2}]{Jiangyan}{Yi}
\name[affiliation={2}]{Tao}{Wang}
\name[affiliation={1,2}]{Yong}{Ren}
\name[affiliation={2}]{Rongxiu}{Zhong}
\name[affiliation={4}]{Zhengqi}{Wen}
\name[affiliation={3}]{Jianhua}{Tao}

\address{
    \textsuperscript{1}School of Artificial Intelligence, University of Chinese Academy of Sciences, Beijing, China \\
    \textsuperscript{2}Institute of Automation,
    Chinese Academy of Sciences, Beijing, China \\
    \textsuperscript{3}Department of Automation, Tsinghua University, Beijing, China \\
    \textsuperscript{4}Qiyuan Laboratory, Beijing, China 
}
\email{le.xu@nlpr.ia.ac.cn, jiangyan.yi@nlpr.ia.ac.cn}


\begin{document}

\maketitle

\begin{abstract}
Recently, there have been significant advancements in voice conversion, resulting in high-quality performance. However, there are still two critical challenges in this field. Firstly, current voice conversion methods have limited robustness when encountering unseen speakers. Secondly, they also have limited ability to control timbre representation. To address these challenges, this paper presents a novel approach that leverages tokens of multi-layer residual approximations to enhance robustness when dealing with unseen speakers, called the residual speaker module. Introducing multi-layer approximations facilitates the separation of information from the timbre, enabling effective control over timbre in voice conversion. The proposed method outperforms baselines in subjective and objective evaluations, demonstrating superior performance and increased robustness. Our demo page is publicly available\footnote{https://frostmiku.github.io/rsm}.
\end{abstract}

\begin{figure*}[htbp]
    \centering
    \includegraphics[height=0.39\linewidth]{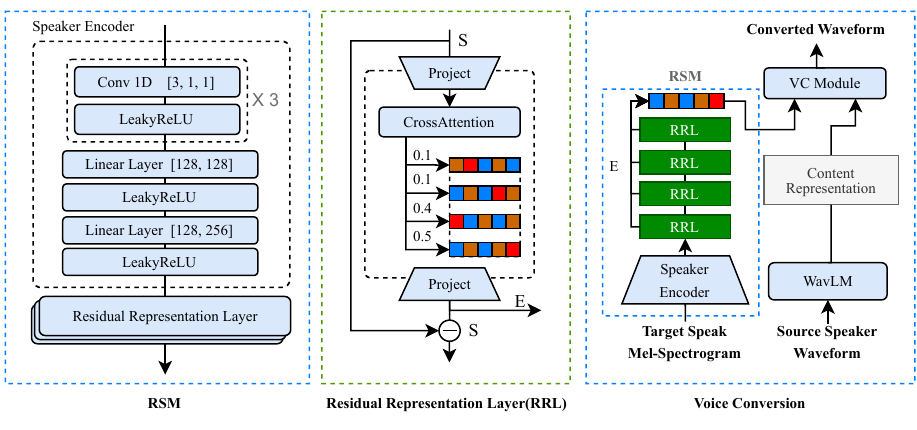}
    \caption{Framework of the voice conversion and speaker representation control}
    \label{fig:framework}
\end{figure*}

\section{Introduction}

Voice conversion seeks to modify different voice attributes, including emotions \cite{zhou2022emotional}, prosody \cite{csicsman2017prosody}, and speaker identity \cite{mohammadi2017overview}, while maintaining the inherent semantic content of the voice. This study focuses on converting speaker identity in one-shot scenarios. Recent advancements in voice conversion have led to remarkable achievements, generating high-quality audio that is becoming more and more similar to natural speech \cite{zhao2020voice, huang2023singing, casanova2022yourtts}. 


Two critical challenges still exist in the field of voice conversion. Firstly, traditional voice conversion methods perform exceptionally well in converting voices when the target speakers are known \cite{liu2021any, huang2021pretraining, kim2021conditional} or depend on a pretrained speaker encoder \cite{sisman2020overview,qian2020f0,walczyna2023overview}. However, they often struggle when faced with out-of-distribution (OOD) caused by previously unseen speakers \cite{li2023freevc, wang2021vqmivc, lin2021fragmentvc}. This insufficient robustness to unseen speakers remains a significant challenge as practical applications frequently require the ability to perform voice conversion for speakers not present in the training dataset. Secondly, most existing voice conversion methods have insufficient control over the timbre attributes \cite{liu2021any,li2023freevc,wang2021vqmivc}, making it still a challenging task to adjust the timbre details while maintaining the identity of the target speaker.

Recent studies \cite{li2023freevc, qian2019autovc,xiao2022dgc,wang2023alo,lian2022robust} have employed pre-trained speaker representation models for encoding timbre representations. The powerful generalization ability of these models depends on the diversity of data in the pre-training phase, while these representations often incorporate extraneous information, such as language and accents, which may compromise the effectiveness of voice conversion models. Global Style Tokens (GST) \cite{wang2018gst} were proposed for global style control in Text-to-Speech tasks. Reference \cite{zhang2021one, xiao2022dgc, lu2019gse, wang2020one} applied GST to speaker representations in voice conversion tasks, while \cite{zhang2021one} used GST combine speaker representations from pre-extracted X-vectors \cite{snyder2018x}, and \cite{xiao2022dgc} used similar methods with D-vector \cite{variani2014deep}. These methods employ learnable tokens to represent the speaker, thus partially mitigating the OOD issue encountered with unseen speakers. Researchers have the ability to modify tokens to alter the voice. However, the accuracy of this approximate representation and the level of control over these modifications is limited.

In response to these challenges, this paper introduces the Residual Speaker Module (RSM). This innovative approach addresses the aforementioned issues by employing tokens of multi-layer approximation techniques to enhance robustness when handling previously unseen speakers. Specifically, during the training phase, a specialized attention mechanism is utilized to map the speaker's voice, which is extracted by the speaker encoder, into multiple sets of trainable tokens. The tokens are included into layers with residual connections, where each layer captures the residual information exclusively from the preceding layer. This can be considered as modeling deviation layer by layer. In the inference phase, the unseen speaker is represented through the combination of these tokens, thereby alleviating the OOD issues and improving the robustness of the model. In addition, the hierarchical residual structure enables a more precise representation of the speaker, enhancing the similarity of the converted audio. Furthermore, it provides researchers with finer control over the voice through modification of each layer of tokens.

We compared the VC system implemented using the RSM method with several baselines on the VCTK \cite{veaux2017cstr} and LibriTTS \cite{zen2019libritts} dataset. Our approach significantly improved system performance and speaker similarity in subjective and objective evaluations. The effectiveness of the method was confirmed through our ablation experiments. Additionally, we investigated the implementation of voice control.

Our contributions include the following: 
(\romannumeral 1) We propose the Residual Speaker Module, which enhances the robustness of the voice conversion model to handle unseen speakers during the inference phase. 
(\romannumeral 2) Our method of layer-wise error modeling has enhanced performance.
(\romannumeral 3) We achieve a degree of control over voice attributes.

\section{Method}

As illustrated in Figure \ref{fig:framework}, our VC system is built on FreeVC \cite{li2023freevc}, which is a conditional VAE \cite{Kingma2013AutoEncodingVB} model based on VITS \cite{kim2021conditional}. We describe RSM's details and elucidate its integration within the pipeline for controllable voice conversion. 

\subsection{Residual Speaker Module}

The residual speaker module is used to compress a variable-length audio signal into a ﬁxed-length vector. As illustrated in Figure \ref{fig:framework}, the proposed Residual Speaker Module primarily comprises two components: (1) A speaker encoder is composed of convolutions and linear layers. (2) A residual representation layer consisting of CrossAttention and residual connections.


\begin{figure}[tbp]
    \centering
    \includegraphics[width=\linewidth]{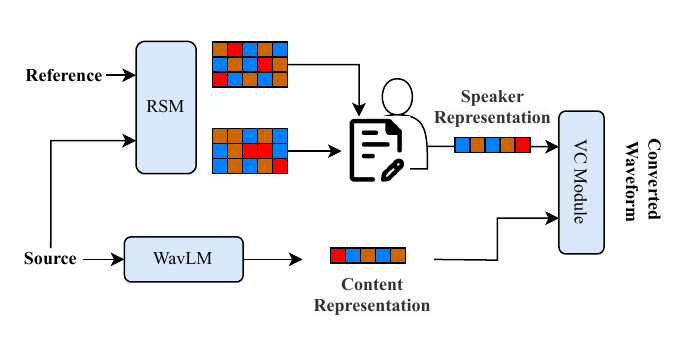}
    \caption{Pipeline of the voice control}
    \label{fig:ve}
\end{figure}

\subsubsection{Speaker Encoder}

Given that speaker representations can be viewed as time-invariant expressions that solely depend on the intrinsic features of the speaker, utilizing a fixed-length vector makes the representation less susceptible to temporal variance. Besides, source and target utterances have different lengths; the reference embedding can not be extracted at the frame level. Hence, we employ a fixed-length speaker embedding method. The speaker encoder is tasked with extracting fixed-length speaker representation vectors $S$ from utterances. Its structure is depicted in Figure \ref{fig:framework}.

Specifically, we employ temporal convolution and linear layers to extract a fixed-length vector from each frame of the mel-spectrogram of the speech signal. Subsequently, we compute the mean value along each dimension. Assuming the input is $X=[x_1, x_2, x_3, \cdots, x_T]$, the definition of the speaker encoder is as follows:
\begin{equation}
    S=\frac{\sum^T_i SpeakerEncoder(x_i)}{T}
\end{equation}
where $S\in \mathbb{R}^{1\times d_s}$, $T$ is the number of frames in the input.

\subsubsection{Residual Representation Layer}

We apply a residual representation layer to constrain the output of the speaker encoder, as illustrated in Figure \ref{fig:framework}. 
We project S to a $\frac{d_s}{\alpha}$-dimensional space through a linear layer. $\alpha$ represents a hyperparameter, which we set to 4 in our study. Intuitively, this operation will retain the primary information in the speaker representation while filtering out secondary details to optimize the final loss. Subsequently, we transform $S$ into a token combination representation within a learnable codebook.

Specifically, we employ $n$ learnable $1 \times \frac{d_s}{\alpha}$-dimensional randomly initialized vectors as token. Then, we combine these vectors into a matrix $C$, which serves as the key and value for the CrossAttention mechanism. we use $S$ as the query for the CrossAttention to compute the speaker embedding $E$ and project it into $\mathbb{R}^{1 \times d_s}$. Mathematically, this can be expressed as follows:
\begin{equation}
    E = softmax(\frac{(SW_q)(CW_k)^T}{\sqrt{d_s}}) \times CW_v \times W_o
\end{equation}

where $d_s$ is the dimension of $S$. $C\in \mathbb{R}^{n\times \frac{d_s}{\alpha}}$ and $E\in \mathbb{R}^{1\times d_s}$. $W_o$ is a matrix of dimensions $\frac{d_s}{\alpha} \times d_s$, whereas $W_q$, $W_k$, and $W_v$ are all $\frac{d_s}{\alpha} \times \frac{d_s}{\alpha}$ matrices.

The process is similar to the GST, which can be viewed as a soft clustering method or an approximation for representing speakers using $n$ factor vectors. A smaller value of $n$ would greatly limit the approximation capability of the RSM module for representing speakers. Hence, we adopt a multi-layer approximation approach based on residual connections to model errors. 
Specifically, for $K$ layers of CrossAttention $A=[A_1, A_2, \cdots, A_k]$, we perform a subtraction operation on the $S$ and $E$ of the $A_1$ layer, and use the residual as the query for the $A_2$ layer. Finally, we sum up the results from all $K$ layers as the final output. The overall computational procedure is depicted in Algorithm \ref{alg:algorithm}.

\begin{algorithm}[tbp]
    \caption{Residual Speaker Module Algorithm}
    \label{alg:algorithm}
    \textbf{Input}: Mel-spectrograms $x_{mel}$ \\
    \textbf{Parameter}: $K$ layers CrossAttention $A_{1 \cdots k}$ with $C_{1 \cdots k}$ \\
    \textbf{Output}: Speaker embedding $E$ 
      \begin{algorithmic}[1] 
      \STATE Let $E=0.0$
      \STATE Extract $S$ from $x_{mel}$
      \FOR {$i=0$ to $K$}
      \STATE $E \leftarrow A_i(S,C_i)+E$
      \STATE $S \leftarrow S-E$
      \ENDFOR
      \STATE \textbf{return} $E$
      \end{algorithmic}
\end{algorithm}

\subsection{Voice Conversion}
As the inference phase, Figure \ref{fig:framework} illustrates the pipeline of voice conversion. We employ the RSM to compute the target speaker's timbre representation from the mel-spectrogram serves as a condition to input to the VC modules. VC module is trained to synthesize speech from given timbre representation and linguistic content.



By employing the residual representation layer for layer-wise error modeling, as the number of codebook layers increases, the influence of the later layers on the final timbre gradually diminishes. Consequently, we can achieve partly voice control. As the Figure\ref{fig:ve}, the ability to selectively adjust token weights in the final layer while preserving the integrity of preceding layers empowers us to create a synthesized speech that retains a desired resemblance to the reference while introducing subtle variations. Alternatively, adjusting tokens in the earlier layers can result in more substantial changes to the timbre. This flexibility in timbre manipulation proves invaluable for applications requiring personalized voice synthesis or subtle modifications.  It is noteworthy that the content encoded by tokens is hyperparameter-dependent, but remains fixed during the inference phase.

\section{Experiments}

\subsection{Datasets}
We conducted experiments on the VCTK \cite{veaux2017cstr} and LibriTTS \cite{zen2019libritts} datasets. Only the VCTK dataset is used in the training phase, which means all evaluations on the LibriTTS dataset are conducted under unseen scenarios for the model. All audio samples are downsampled to 16 kHz, and then audio normalization is applied to them. Mel-spectrograms are calculated using a short-time Fourier transform. The FFT, window, and hop sizes are set to 1280, 1280, and 320, respectively.

\subsection{Implementation Details}
Our models and backbone are trained up to 350k steps on a single NVIDIA A100 GPU. The batch size is set to 64 with a maximum segment length of 128 frames. We use the AdamW optimizer\cite{loshchilov2017decoupled} and set $\beta_1$ = 0.8, $\beta_2$ = 0.99 and weight decay $\lambda$ = 0.01. We use the Exponential learning rate decay scheduler with a 0.999875 factor in every epoch, where the initial learning rate is set to 0.0002. The seed of the random number generator is set to 1234. We adopt slice training, a method of using only a part of frames for calculating loss, to reduce training time and memory usage during training. All baselines use the same settings.

\subsection{Baseline}
As described in Table \ref{tab:baseline}, we selected three speaker representation methods to apply as baselines compared with our model. GT is the ground truth. B01-B03 are baseline systems. B01 and B02 were proposed by FreeVC and we used the same setup as the original method. In B03, the speaker encoder was replaced with a jointly trained GST. P01 is our proposed method, and P02 is an ablation study, which is the same as P01 but without residual connections.

\subsection{Evaluation Metrics}

We use the open-source ASR system \footnote{https://huggingface.co/facebook/hubert-large-ls960-ft} to test the Character Error Rate (CER) and Word Error Rate (WER) of the converted utterance to evaluate whether the converted utterance maintains the linguistic content and intonation variations of the source utterance. Note that the linguistic content is unseen during the training phase in our evaluations.

For subjective evaluation, we employ the Mean Opinion Score (MOS) as our testing standard. The listener needs to give a score for each sample in a test case according to the criterion: 1 = Bad; 2 = Poor; 3 = Fair; 4 = Good; 5 = Excellent \cite{jia2018transfer}. We selected utterances from 10 speakers in both the VCTK and LibriTTS datasets for voice conversion. For each testing session, we randomly extracted 10 samples from the converted audio of each speaker to form the test set. 10 participants were invited to conduct tests on naturalness and speaker similarity, with the results labeled as MOS and SMOS, respectively. We conducted tests separately in unseen scenarios. This means the target speaker was unseen during training.

In terms of Speech Naturalness, B02, B03, and P02, which employ a jointly trained speaker encoder, exhibit similar and lower scores compared to B01, based on a pre-trained speaker encoder. We attribute this to the pre-trained speaker encoder being trained on a large-scale speech dataset, enabling it to better handle unseen scenarios. Our proposed method, P01, achieves scores similar to B01, validating the effectiveness of mitigating OOD issues by transforming speaker representations into known token combinations. This suggests that the approach of P01, through converting speaker representations into known token combinations, is effective in addressing OOD challenges.

\begin{figure*}[htbp]
    \centering
    \includegraphics[trim= 0cm 0cm 0cm 1.7cm, clip, width=\linewidth]{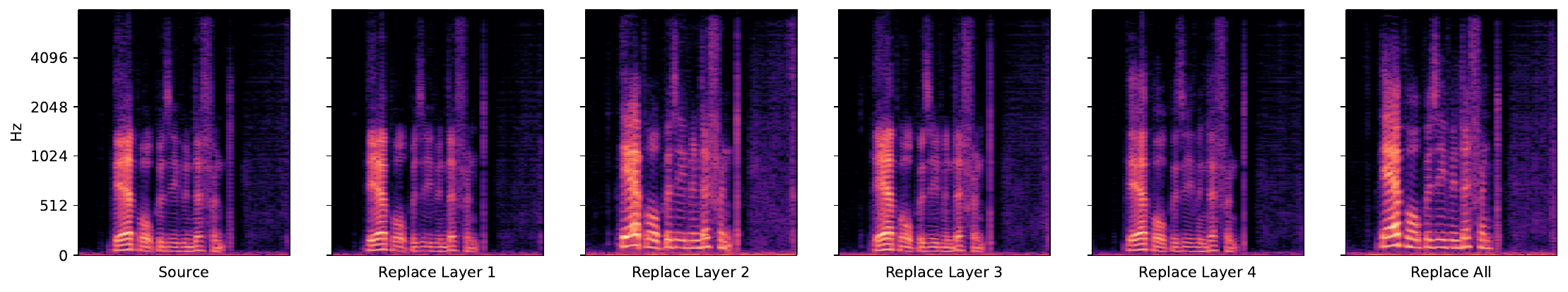}
    \caption{Mel-spectrogram of synthesized speech after replacing speaker representations extracted by RSM layer by layer}
    \label{fig:layer-replacement}
\end{figure*}

\subsection{Results and Discussion}

Our objective and subjective experimental results are presented in Table \ref{tab:result}.

\subsubsection{Objective Evaluation}
For the objective evaluation, the jointly trained speaker representation module contributes to lower WER and CER, which is consistent with findings in the FreeVC paper. P01 achieved the lowest WER and CER, which we attribute to the influence of joint training and token representations. The use of token combinations, to some extent, mitigated the OOD issue. Concurrently, we observed a higher word error rate in P02, which we attribute to the error introduced by the discretized representation of tokens. This observation validates the effectiveness of the multi-layer error modeling approach based on residual representations.

\subsubsection{Subjective Evaluation}

For the subjective, our system exhibits higher speech similarity when dealing with unseen target speakers. All experimental metrics outperform the baselines.

\begin{table}[tbp]
    \caption{Description of systems}
    \begin{tabularx}{\linewidth}[htbp]{cX}
        \toprule
        ID & Describe\\
        \midrule
        GT & Ground truth. \\
        B01 & FreeVC, proposed by \cite{li2023freevc}. \\
        & A pretrained speaker encoder \cite{liu2021any} is used.\\
        B02 & FreeVC-s, proposed by \cite{li2023freevc}. \\
        &A jointly trained speaker encoder is used.\\
        B03 & Replace speaker encoder in FreeVC with GST.\\
        \midrule
        P01 & Replace speaker encoder in FreeVC with 4 layers RSM. \\
        P02 & Ablation study for P01, RSM without residual connections.\\
        \bottomrule
    \end{tabularx}
    \label{tab:baseline}
\end{table}

\setlength{\tabcolsep}{1.8mm}{
\begin{table}[tbp]
    \centering
    \caption{Objective and subjective evaluations. B01-B03 are baselines. P01 and P02 are our proposed methods. MOS and SMOS with 95\% conﬁdence intervals are reported. }
    \begin{tabular}{c|cc|cc}
    \toprule[1pt]
        ID & CER(\%)$\downarrow$ & WER(\%)$\downarrow$ & MOS$\uparrow$ & SMOS$\uparrow$ \\
    \midrule[0.5pt]
        GT & 1.30 & 4.67  & - & - \\
        B01 & 5.62  & 13.17 & 3.82 $\pm$ 0.08 & 3.02 $\pm$ 0.09\\
        B02 & 5.45  & 12.99 & 3.20 $\pm$ 0.09 & 2.96 $\pm$ 0.11\\ 
        B03 & 6.36  & 12.91 & 3.23 $\pm$ 0.11 & 2.59 $\pm$ 0.10\\
    \midrule[0.5pt]
        P01 & \textbf{5.15\%}  & \textbf{11.52\%}  & \textbf{3.85 $\pm$ 0.11} & \textbf{3.46 $\pm$ 0.10}\\ 
        P02 & 6.87\%  & 14.71\% & 3.31 $\pm$ 0.09 & 2.86 $\pm$ 0.13\\ 
    \bottomrule[1pt]
    \end{tabular}
    \label{tab:result}
\end{table}
}
For the Speaker Similarity, GST exhibits the poorest performance, possibly due to limitations imposed by the codebook size, affecting the descriptive capacity of speaker representations. The method proposed in this paper significantly outperforms other approaches, albeit with a wider confidence interval. Considering potential influences from volunteer personal preferences, our demo is available on the webpage\textsuperscript{1}.



\begin{table}[tp]
    \centering
    \caption{The standard deviation of each level}
    \vspace{0.1cm}
    \begin{tabular}{l|cccc}
        \toprule[1pt]
        ID & Layer 1 & Layer 2 & Layer 3 & Layer 4 \\
        \midrule[0.5pt]
        P01 & 0.4835 & 0.4264 & 0.4271 & 0.3995 \\
        P02 & 0.4594 & 0.4626 & 0.4455 & 0.4396  \\
        \bottomrule[1pt]
    \end{tabular}
    \label{tab:std}
\end{table}

\begin{figure}[htp]
    \centering
    \includegraphics[width=\linewidth]{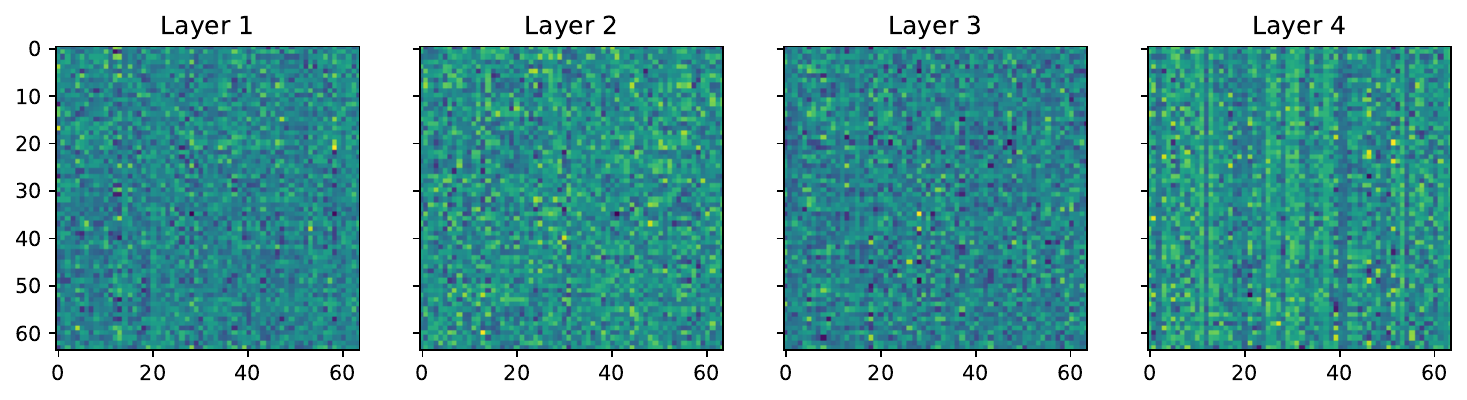}
    \includegraphics[trim= 2.7cm 2cm 2.7cm 2cm, clip, width=\linewidth]{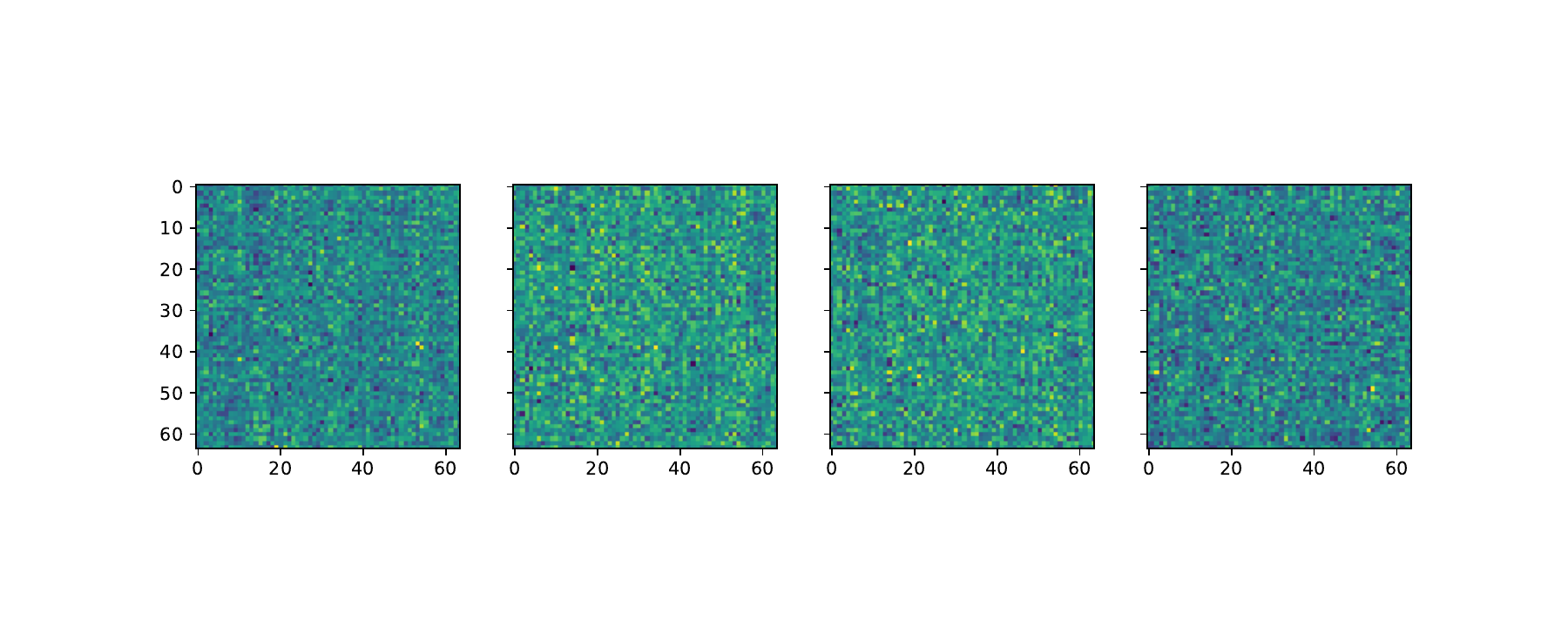}
    \caption{Visualization of the codebook of RSM (frist line) and ablation study (second line)}
    \label{fig:codebook}
\end{figure}


\subsection{Voice Control}

For voice control, we attempted to control tokens at each layer while keeping the other layers fixed when synthesizing speech. Our audio samples are publicly available\textsuperscript{1}.

Figure \ref{fig:layer-replacement} displays the mel-spectrogram of the converted audio when each layer's tokens of the source speaker (male) are individually replaced with corresponding layer tokens of the target speaker (female). 

Participants easily discerned differences in the converted audio compared to the source audio when tokens in the first or second layers were replaced. For the third layer, participants perceived distinctions from the source audio but considered it to originate from the same speaker. When replacing tokens in the fourth layer alone, only a small portion of native speakers could detect the differences.

We quantitatively and qualitatively analyzed the codebook. We computed the mean of standard deviation across dimensions for each layer in P01 and P02, which reflects the amount of information in the codebook, and the results are presented in Table \ref{tab:std}. The standard deviation decreases layer by layer, which can be attributed to the continuous reduction of remaining information. Figure \ref{fig:codebook} illustrates the visualization of P01 and P02, revealing noticeable stripes in the fourth layer of P01, absent in P02 as it serves as an ablation experiment. This observation further confirms the continuous reduction of residual information, indicating a weaker impact of the fourth-layer codebook on the final synthesized speech.

\section{Conclusions}
In this work, we propose the RSM employ tokens of multi-layer approximation and error modeling techniques to enhance robustness when handling previously unseen speakers and provide partial control over voice characteristics. We apply it to build a more robust VC system. Subjective and objective experiments confirm the effectiveness of the proposed module. In future research, we plan to explore finer-grained control over voice attributes.

\section{Acknowledgements}
This work is supported by the Strategic Priority Research Program of Chinese Academy of Sciences, Grant No. XDB0500103, the National Natural Science Foundation of China (NSFC) (No. 62322120, No.U21B2010, No. 62306316, No. 62206278).



\bibliographystyle{IEEEtran}
\bibliography{mybib}

\end{document}